\documentclass[twocolumn,showpacs,aps,pre]{revtex4}

\usepackage{amssymb,amsmath}
\usepackage[dvips]{graphicx}
\usepackage[footnotes,writekey=false]{notes2bib}

\usepackage{hyperref}
\hypersetup{
breaklinks=true,
colorlinks=true,
citecolor=blue,
linkcolor=blue,
filecolor=blue,
urlcolor=blue
}

\begin{document}

\title{Absorbing-state phase transitions on percolating lattices}

\author{Man Young Lee}
%\email{vojtat@mst.edu}
\affiliation{Department of Physics, Missouri University of Science and Technology,
Rolla, MO 65409, USA}
\author{Thomas Vojta}
\affiliation{Department of Physics, Missouri University of Science and Technology, Rolla, MO 65409, USA}
\affiliation{Max-Planck-Institute for Physics of Complex Systems, Noethnitzer Str. 38, 01187
Dresden, Germany}

\begin{abstract}
We study nonequilibrium phase transitions of reaction-diffusion systems defined on
randomly diluted lattices, focusing on the transition across the lattice percolation
threshold. To develop a theory for this transition, we combine classical percolation
theory with the properties of the supercritical nonequilibrium system on a finite-size
cluster. In the case of the contact process, the interplay between geometric criticality
due to percolation and dynamical fluctuations of the nonequilibrium system leads to a new
universality class. The critical point is characterized by ultraslow activated dynamical
scaling and accompanied by strong Griffiths singularities. To confirm the universality of
this exotic scaling scenario we also study the generalized contact process with several
(symmetric) absorbing states, and we support our theory by extensive Monte-Carlo
simulations.
\end{abstract}

\date{\today}

\maketitle

%%%%%%%%%%%%%%%%%%%%%%%%%%%%%%%%%%%%%%%%%%%%%%%%%%%%%%%%%%%%%%%%%%%%%%%%%%%%%%%%%
% Main text starts here
%%%%%%%%%%%%%%%%%%%%%%%%%%%%%%%%%%%%%%%%%%%%%%%%%%%%%%%%%%%%%%%%%%%%%%%%%%%%%%%%%
\section{Introduction}
%%%%%%%%%%%%%%%%%%%%%%%%%%%%%%%%%%%%%%%%%%%%%%%%%%%%%%%%%%%%%%%%%%%%%%%%%%%%%%%%%

In recent years, considerable effort has been directed towards identifying and classifying
phase transitions far from thermal equilibrium. Such nonequilibrium transitions can
be found in a wide variety of problems in biology, chemistry, and physics. Examples
include population dynamics, the spreading of epidemics, surface chemical reactions, catalysis,
granular flow, traffic jams as well as growing surfaces and interfaces (see, e.g.,
\cite{Liggett85,ZhdanovKasemo94,SchmittmannZia95,MarroDickman99,Hinrichsen00,Odor04,Luebeck04,TauberHowardVollmayrLee05}).
Nonequilibrium phase transitions are characterized by large scale fluctuations
and collective behavior in space and time very similar to the behavior at equilibrium
critical points.

A particularly interesting situation arises when an equilibrium or nonequilibrium
many-particle system is defined on a randomly diluted lattice. Then, two distinct types of
fluctuations are combined, \emph{viz.} the dynamical fluctuations of the
many-particle system and the static geometric fluctuations due to lattice percolation
\cite{StaufferAharony_book91}. In
equilibrium systems, their interplay gives rise to novel universality classes for
the thermal \cite{Bergstresser77,StephenGrest77,GefenMandelbrotAharony80}
and quantum \cite{SenthilSachdev96,Sandvik02,VojtaSchmalian05b,WangSandvik06} phase transitions
across the lattice percolation threshold.

In this paper, we investigate the interplay between dynamical fluctuations and geometric
criticality
in nonequilibrium many-particle systems. We focus on a particularly well-studied type of
transitions, the so-called absorbing state transitions, that separate active, fluctuating steady
states from inactive (absorbing) states in which fluctuations cease completely.
The generic universality class for absorbing state transitions is directed percolation
(DP)
\cite{GrassbergerdelaTorre79}. It is conjectured \cite{Janssen81,Grassberger82} to be valid
for all absorbing state transitions with scalar order parameter and no extra symmetries
or conservation laws. In the presence of symmetries and/or conservation laws, other universality
classes can be realized, such as the DP$n$ class in systems with $n$ symmetric absorbing states
\cite{Hinrichsen97}.

For definiteness, we consider the contact process \cite{HarrisTE74}, a prototypical system
in the DP universality class. We show that the contact process on a randomly site or bond
diluted lattice has two different nonequilibrium phase transitions: (i) a generic disordered
DP transition at weak dilutions (below the lattice percolation threshold) driven by the dynamic
fluctuations of the contact process and (ii) the
transition across the lattice percolation threshold driven by the geometric criticality of
the lattice.
The former transition has been investigated for a number of years
\cite{Kinzel85,Noest86,MoreiraDickman96,Janssen97}; it has recently reattracted considerable attention because
it is governed by an exotic infinite-randomness fixed point
\cite{HooyberghsIgloiVanderzande03,HooyberghsIgloiVanderzande04,VojtaDickison05,VojtaFarquharMast09}.
In contrast, the latter transition has received much less attention.

Here, we develop a theory for the nonequilibrium transition across the lattice
percolation threshold by combining classical percolation theory with the properties of
the supercritical contact process on a finite-size cluster. We show that the critical
point is characterized by ultraslow activated (exponential) dynamical scaling and
accompanied by strong Griffiths singularities. The scaling scenario is qualitatively
similar to the generic disordered DP transition, but with different critical exponent
values. To confirm the universality of this exotic scenario, we also investigate the
generalized contact process with $n$ (symmetric) absorbing states \cite{Hinrichsen97}.
This is a particularly interesting problem because the generic transition of the
disordered generalized contact process does \emph{not} appear to be of
infinite-randomness type
\cite{HooyberghsIgloiVanderzande03,HooyberghsIgloiVanderzande04}.

The paper is organized as follows. In Sec.\ \ref{sec:processes}, we introduce our
models, the simple and generalized contact processes on a randomly diluted lattice.
We also discuss the phase diagrams. In Sec.\ \ref{sec:percolation} we briefly summarize
the results of classical percolation theory to the extent necessary for our purposes.
Section \ref{sec:theory} contains the main part of the paper, the theory of the
nonequilibrium transition across the lattice percolation threshold. Section
\ref{sec:generality} is devoted to the question of the generality of the arising
scaling scenario. We conclude in Sec.\ \ref{sec:conclusions}. A short account of part of
this work has already been published in Ref. \cite{VojtaLee06}.

%%%%%%%%%%%%%%%%%%%%%%%%%%%%%%%%%%%%%%%%%%%%%%%%%%%%%%%%%%%%%%%%%%%%%%%%%%%%%%%%%
\section{Simple and generalized contact processes on diluted lattices}
\label{sec:processes}
%%%%%%%%%%%%%%%%%%%%%%%%%%%%%%%%%%%%%%%%%%%%%%%%%%%%%%%%%%%%%%%%%%%%%%%%%%%%%%%%%

\subsection{Contact process}

The clean contact process \cite{HarrisTE74} is a prototypical system in the DP
universality class. It is defined on a $d$-dimensional hypercubic lattice. (We consider
$d\ge2$ since we will be interested in diluting the lattice.) Each lattice site
$\mathbf{r}$ can be active (infected, state A) or inactive (healthy, state I).
During the time evolution of the contact process which is a continuous-time Markov
process, each active site becomes inactive at a rate $\mu$ (``healing'') while each inactive
site becomes active at a rate $\lambda m / (2d)$ where $m$ is the number of active nearest
neighbor sites (``infection''). The infection rate
$\lambda$ and the healing rate $\mu$
are external parameters. Their ratio controls the behavior of the contact process.

For $\lambda \ll \mu$, healing dominates over infection, and the absorbing state without any active
sites is the only steady state of the system (inactive phase). For sufficiently large infection rate
$\lambda$, there is a steady state with a nonzero density of active sites (active phase).
These two phases are separated by a nonequilibrium phase transition in the DP universality
class at a critical value $(\lambda/\mu)_c^0$ of the ratio of the infection and healing
rates.

The basic observable in the contact process is the average density of active sites
at time $t$,
\begin{equation}
\rho(t) = \frac 1 {L^d} \sum_{\mathbf{r}} \langle  n_\mathbf{r}(t) \rangle
\label{eq:rho_definition}
\end{equation}
where $n_\mathbf{r}(t)=1$ if the site $\mathbf{r}$ is active at time $t$ and $n_\mathbf{r}(t)=0$
if it is inactive. $L$ is the linear system size, and $\langle \ldots \rangle$ denotes the
average over all
realizations of the Markov process. The longtime limit of this density (i.e., the steady
state density)
\begin{equation}
\rho_{\rm stat} = \lim_{t\to\infty} \rho(t)
\label{eq:OP_definition}
\end{equation}
is the order parameter of the nonequilibrium phase transition.

\subsection{Generalized contact process}

Following Hinrichsen \cite{Hinrichsen97}, we now generalize the contact process
by introducing $n$ different inactive states I$_k$ with $k=1 \ldots n$ ($n=1$
corresponds to the simple contact process). Here,
$k$ is sometimes called the ``color'' label. The time evolution is
again a continuous-time Markov process. The first two rates are equivalent to those of
the simple contact process: An active site can decay into each of the inactive states
I$_k$ with rate $\mu/n$, and a site in any of the inactive states becomes active at
a rate $\lambda m / (2d)$ with $m$ the number of active nearest-neighbor sites.
To introduce competition between the different inactive states, we define a third rate:
If two neighboring sites are in \emph{different} inactive states, each can become
active with a rate $\sigma$. This last rule prevents the boundaries between domains
of different inactive states from sticking together infinitely. Instead they can
separate, leaving active sites behind.

The properties of the clean generalized contact process have been studied in some detail
in the literature \cite{Hinrichsen97,HooyberghsCarlonVanderzande01}. If the boundary
activation rate $\sigma$ vanishes, the behavior becomes identical to the simple contact
process for all $n$. (This becomes obvious by simply dropping the color label and treating all
inactive sites as identical.) For $\sigma>0$, the system becomes ``more active'' than the
simple contact process, and the universality class changes. In one space dimension, a
phase transition exists for $n=1$ (in the DP universality class) and for $n=2$
(in the $Z_2$-symmetric directed percolation (DP2) class which coincides with the
the parity-conserving (PC) class in one dimension \cite{Hinrichsen00}). For $n\ge 3$ the
system is always in the active phase, and no phase transition exists at finite values of
$\lambda,\mu$ and $\sigma$.

The generalized contact process in higher space dimensions presumably behaves in an
analogous fashion:
There is a DP transition for $n=1$ while the properties for $n>1$ are different.
For sufficiently large $n$, the system is always active
\footnote{For $d=2, n=2$, Hinrichsen \cite{Hinrichsen97} finds a
mean-field transition while our own simulations suggest that the system always
active. Since this difference is of no importance for the present paper,
it will be addressed elsewhere.}.

\subsection{Lattice dilution}

We now introduce quenched site dilution by randomly removing each lattice site with
probability $p$. (Bond dilution could be introduced analogously.)
As long as the vacancy concentration $p$ remains below the lattice percolation
threshold $p_c$, the lattice consists of an infinite connected cluster of sites accompanied
by a spectrum of finite-size clusters. In contrast, at dilutions above $p_c$, the lattice
consists of disconnected finite-size clusters only.

\begin{figure}
\includegraphics[width=7.cm,clip]{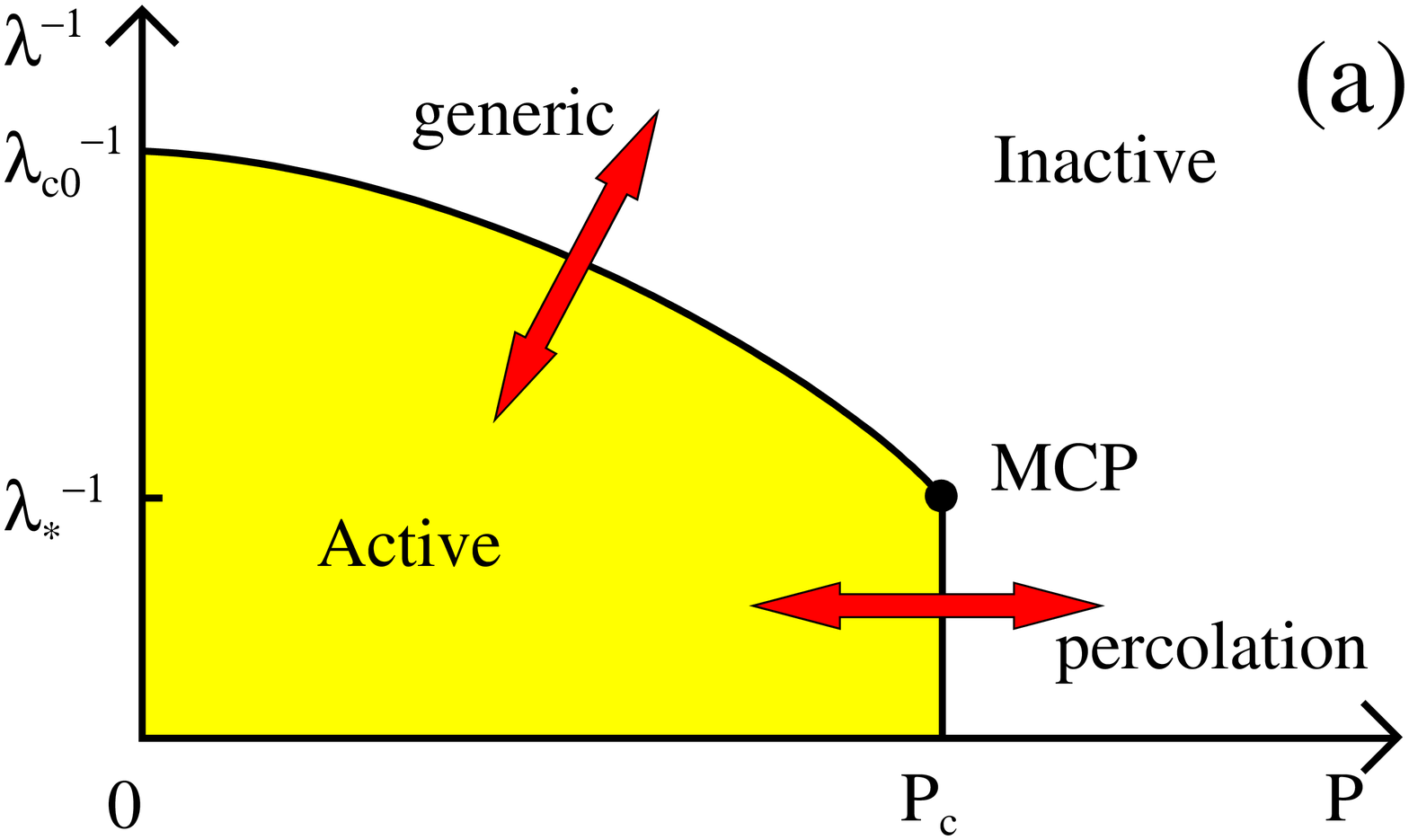}\\[2ex]
\includegraphics[width=7.cm,clip]{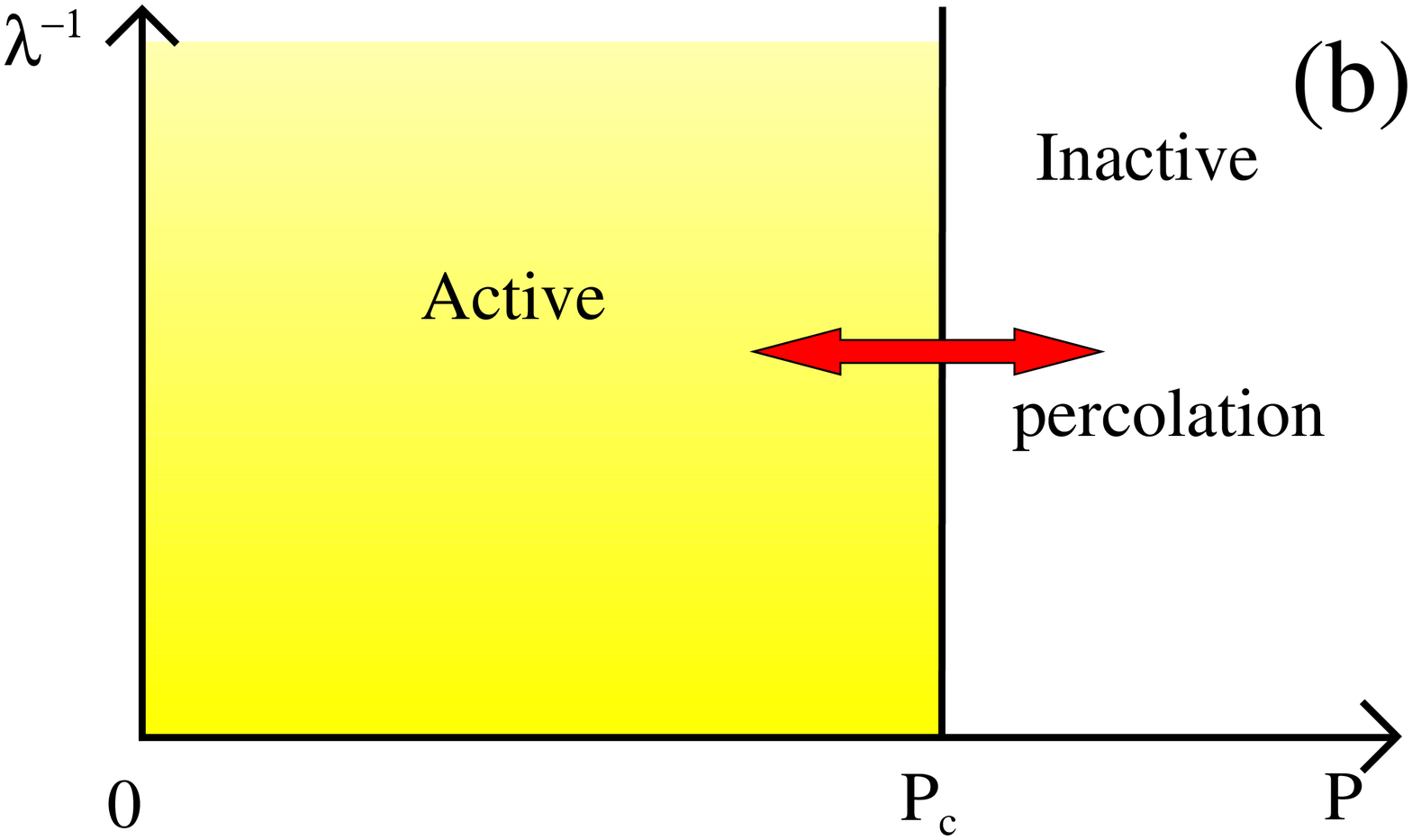}
\caption{(Color online:) Schematic phase diagrams for the simple and generalized contact
processes on a diluted lattice in dimensions $d\ge 2$ as a function of dilution $p$ and
inverse infection rate $\lambda^{-1}$ (healing and boundary activation rates $\mu$ and
$\sigma$ are fixed). Case (a) applies to systems that display a phase transition at
$\lambda_c^0$ in the absence of dilution. There is a multicritical point (MCP) at
$(p_c,\lambda_*)$ separating the generic transition from the lattice percolation
transition. Case (b) is for systems that are always active in the absence of dilution.}
\label{fig:pds}
\end{figure}

Figure \ref{fig:pds} schematically shows the resulting phase diagrams of the
nonequilibrium process as a function of the infection rate $\lambda$ and dilution $p$,
keeping the healing rate $\mu$ and the boundary activation rate $\sigma$, if any,
constant.
Depending on the properties of the clean undiluted system, there are two qualitatively
different cases.

(a) If the undiluted system has a phase transition at a nonzero critical infection rate
$\lambda_c^0$, the active phase survives for all vacancy concentrations below the
percolation threshold, $p<p_c$. It even survives at the percolation threshold $p_c$ on
the critical percolation cluster because it is connected, infinitely extended, and its
fractal dimension $D_f$ is larger than unity. The critical infection rate $\lambda_c$
increases with increasing dilution $p$ to compensate for the missing neighbors, reaching
$\lambda_*$ at $p_c$. The active phase cannot exist for $p>p_c$ because the lattice
consists of finite-size clusters only, and the nonequilibrium process will eventually end
up in one of the absorbing states on any finite-size cluster.  Thus, in case (a), our
system features two nonequilibrium phase transitions, (i) a generic (disordered)
transition for dilutions $p<p_c$, driven by the dynamic fluctuations of the
nonequilibrium process and (ii) the transition across the lattice percolation threshold
driven by the geometric criticality of the lattice. They are separated by a multicritical
point at $(p_c,\lambda_*)$ which was studied numerically in Ref.\ \cite{DahmenSittlerHinrichsen07}.

(b) If the undiluted system is always active (as for the generalized contact
process with a sufficiently high number of inactive states), the phase diagram is
simpler. The active phase covers the entire region $p\le p_c$ for all $\lambda>0$
($\lambda_*$ is formally zero)
while the inactive phase exists in the region $p>p_c$. There is no generic (disordered)
nonequilibrium phase transition, only the transition across the lattice percolation
threshold.

The focus of the present paper is the nonequilibrium phase transition across the
lattice percolation threshold that exists in both cases. In order to develop a theory
for this transition, we combine classical percolation theory with the properties
of the nonequilibrium process on a finite-size cluster. In the next section we therefore
briefly summarize key results of percolation theory.

%%%%%%%%%%%%%%%%%%%%%%%%%%%%%%%%%%%%%%%%%%%%%%%%%%%%%%%%%%%%%%%%%%%%%%%%%%%%%%%%%
\section{Classical percolation theory}
\label{sec:percolation}
%%%%%%%%%%%%%%%%%%%%%%%%%%%%%%%%%%%%%%%%%%%%%%%%%%%%%%%%%%%%%%%%%%%%%%%%%%%%%%%%%

Consider a regular lattice in $d$ dimensions. If each lattice site is removed with
probability $p$ \footnote{We define $p$ as the fraction of sites removed rather than
the fraction of sites present.}, an obvious question is whether or not the lattice is
still connected in the sense that there is a cluster of connected
(nearest neighbor) sites that spans the entire system. This question defines the
percolation problem (see Ref.\ \cite{StaufferAharony_book91} for an introduction).

In the thermodynamic limit of infinite system volume, there is a sharp boundary
between the cases of a connected or disconnected lattice. If the vacancy concentration
$p$ stays below the percolation threshold $p_c$, an infinite cluster of connected sites
exists (with a probability of unity). For $p>p_c$, an infinite cluster does not exist,
instead, the lattice consists of many disconnected finite-size clusters.

The behavior of the lattice for vacancy concentrations close to the percolation threshold
can be understood as a (geometric) continuous phase transition or critical phenomenon.
The order parameter is the probability $P_\infty$ of a site to belong to the infinite
connected percolation cluster. It is obviously zero in the disconnected phase ($p>p_c$)
and nonzero in the percolating phase ($p<p_c$). Close to $p_c$ it varies as
\begin{equation}
P_\infty \sim |p-p_c|^{\beta_c} \qquad (p<p_c)
\label{eq:percolation-beta}
\end{equation}
where $\beta_c$ is the order parameter critical exponent of classical percolation. Note
that we use a subscript $c$ to distinguish quantities associated with the classical
lattice percolation problem from those of the nonequilibrium phase transitions discussed
later. In addition to the infinite cluster, we also need to characterize the finite
clusters on both sides of the transition. Their typical size, the correlation or
connectedness length $\xi_c$ diverges as
\begin{equation}
\xi_c \sim |p-p_c|^{-\nu_c}
\label{eq:percolation-nu}
\end{equation}
with $\nu_c$ the correlation length exponent. The average mass $S_c$ (number of sites) of a
finite cluster diverges with the susceptibility exponent $\gamma_c$ according to
\begin{equation}
S_c \sim |p-p_c|^{-\gamma_c}~.
\label{eq:percolation-gamma}
\end{equation}

The complete information about the percolation critical behavior is contained in the cluster
size distribution $n_s$, i.e., the number of clusters with $s$ sites excluding the infinite
cluster (normalized by the total number of lattice sites). Close to the percolation
threshold, it obeys the scaling form
\begin{equation}
n_{s} ( \Delta) =s^{-\tau_c }f\left( \Delta s^{\sigma_c }\right) .
\label{eq:percscaling}
\end{equation}
Here $\Delta=p-p_c$, and $\tau_c $ and $\sigma_c$ are critical exponents.
The scaling function $f(x)$ is analytic for small $x$ and has a single maximum at
some $x_{\rm max}>0$. For large $|x|$, it drops off rapidly
\begin{eqnarray}
f(x) &\sim&  \exp\left[- B_1 x^{1/\sigma_c}\right] \quad ~(x>0),
\label{eq:scaling-function-disconnected}\\
f(x) &\sim&     \exp\left[- \left(B_2 x^{1/\sigma_c}\right)^{1-1/d}\right] \quad ~(x<0),
\label{eq:scaling-function-connected}
\end{eqnarray}
where $B_1$ and $B_2$ are constants of order unity. All classical percolation exponents are
determined by $\tau_c$ and $\sigma_c$ including the correlation lengths exponent
$\nu_c =({\tau_c -1})/{(d\sigma_c )}$, the order parameter exponent
$\beta_c=(\tau_c-2)/\sigma_c$, and the susceptibility exponent
$\gamma_c=(3-\tau_c)/\sigma_c$.

Right at the percolation threshold, the cluster size distribution does not contain a
characteristic scale. The structure of the critical percolation cluster is thus fractal
with the fractal dimension being given by $D_f=d/(\tau_c-1)$.

%%%%%%%%%%%%%%%%%%%%%%%%%%%%%%%%%%%%%%%%%%%%%%%%%%%%%%%%%%%%%%%%%%%%%%%%%%%%%%%%%
\section{Nonequilibrium transition across the lattice percolation threshold}
\label{sec:theory}
%%%%%%%%%%%%%%%%%%%%%%%%%%%%%%%%%%%%%%%%%%%%%%%%%%%%%%%%%%%%%%%%%%%%%%%%%%%%%%%%%

\subsection{Single-cluster dynamics}
\label{subsec:single-cluster}

To develop a theory of the nonequilibrium phase transition across the lattice percolation
threshold, we first study the nonequilibrium process on a single connected finite-size cluster of
$s$ sites. For definiteness, this section focuses on the simple contact process. The generalized
contact process will be considered in Sec.\ \ref{sec:generality}.

\begin{figure}
\includegraphics[width=6.5cm,clip]{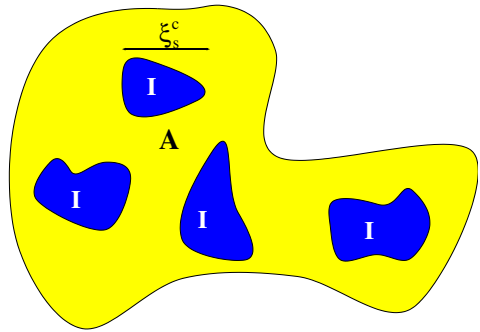}
\caption{(Color online:) Schematic of the metastable state of the supercritical
         contact process on a single percolation cluster.
         A and I denote active and inactive sites, and $\xi_s^c$ is the connected
         correlation length of the density fluctuations \emph{on} the cluster.}
\label{fig:cp_cluster}
\end{figure}

The crucial observation is that on the percolation transition line (for
$\lambda>\lambda_*$), the contact process is supercritical, i.e., the cluster is locally in the active phase.
The time evolution of
such a cluster, starting from a fully active lattice, therefore proceeds in two stages:
Initially, the density $\rho_s$ of active sites decays rapidly towards a metastable state
(which corresponds to the steady state of the equivalent \emph{infinite} system) with a
nonzero density of active sites and islands of the inactive phase of linear size
$\xi_s^{c}$ (see Fig.\ \ref{fig:cp_cluster}).
This
metastable state can then decay into the inactive (absorbing) state only via a rare
collective fluctuation involving \emph{all} sites of the cluster. We thus expect
the long-time decay of the density to be of exponential form (suppressing subleading
pre-exponential factors),
\begin{equation}
\rho_s(t) \sim  \exp[{-t/t_s(s)}]~,
\label{eq:cluster-decay}
\end{equation}
with a long lifetime $t_s$ that increases exponentially with the cluster size $s$
\begin{equation}
t_s(s) = t_0 \exp[{A(\lambda)s}]
\label{eq:cluster-lifetime}
\end{equation}
for sufficiently large $s$. Here, $t_0$ is some microscopic time scale.

The lifetime increases the faster with $s$
the further the cluster is in the active phase. This means, the prefactor $A(\lambda)$ which
plays the role of an inverse correlation volume vanishes
at the multicritical value $\lambda_*$ and monotonically increases with increasing
$\lambda$. Close to the multicritical point, the behavior of $A(\lambda)$ can be inferred
from scaling. Since $A(\lambda)$ has the dimension of an inverse volume, it varies as
\begin{equation}
A(\lambda) \sim (\lambda-\lambda_*)^{\nu_* D_f}
\label{eq:Alambda}
\end{equation}
where $\nu_*$ is the correlation length exponent of the multicritical point and $D_f$ is
the (fractal) space dimensionality of the underlying cluster.

Note that (\ref{eq:cluster-lifetime}) establishes an exponential relation between
length and time scales at the transition. Because the number of sites $s$ of a
percolation cluster is related to its linear size $R_s$ via $s \sim R_s^{D_f}$,
eq.\ (\ref{eq:cluster-lifetime}) implies
\begin{equation}
\ln t_s \sim R_s^{D_f}~.
\label{eq:activated-scaling}
\end{equation}
Thus, the dynamical scaling is activated rather than power-law with the tunneling
exponent being identical to the fractal dimension of the critical percolation cluster,
$\psi=D_f$.

To confirm the above phenomenological arguments, we have performed extensive Monte-Carlo
simulations of the contact process on finite-size clusters using clean one-dimensional
and two-dimensional systems as well as diluted lattices. Our simulation method is based on the
algorithm by Dickman \cite{Dickman99} and described in detail in Refs.\
\cite{VojtaDickison05,VojtaFarquharMast09}.

\begin{figure}
\includegraphics[width=8cm,clip]{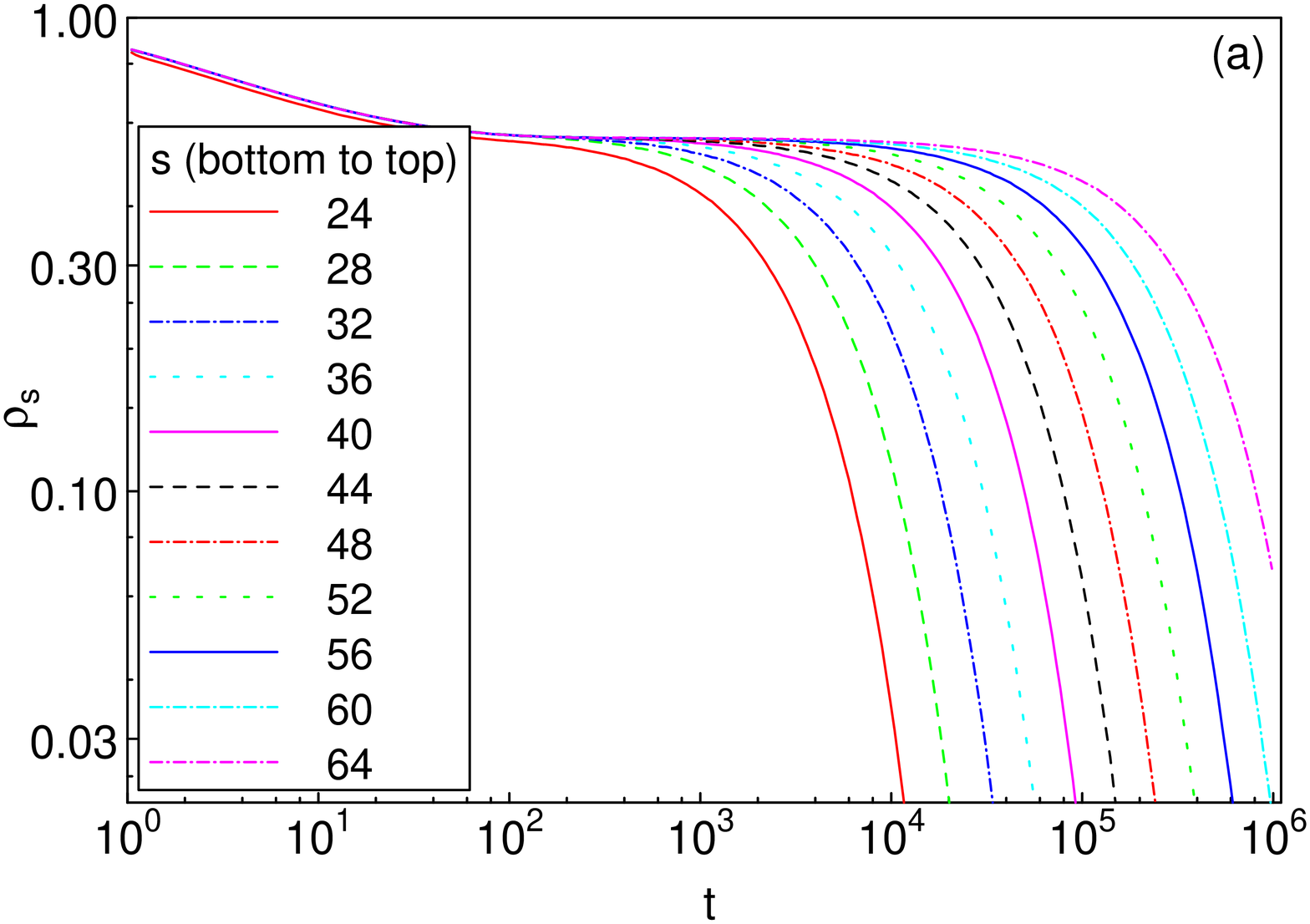}\\[2ex]
\includegraphics[width=8cm,clip]{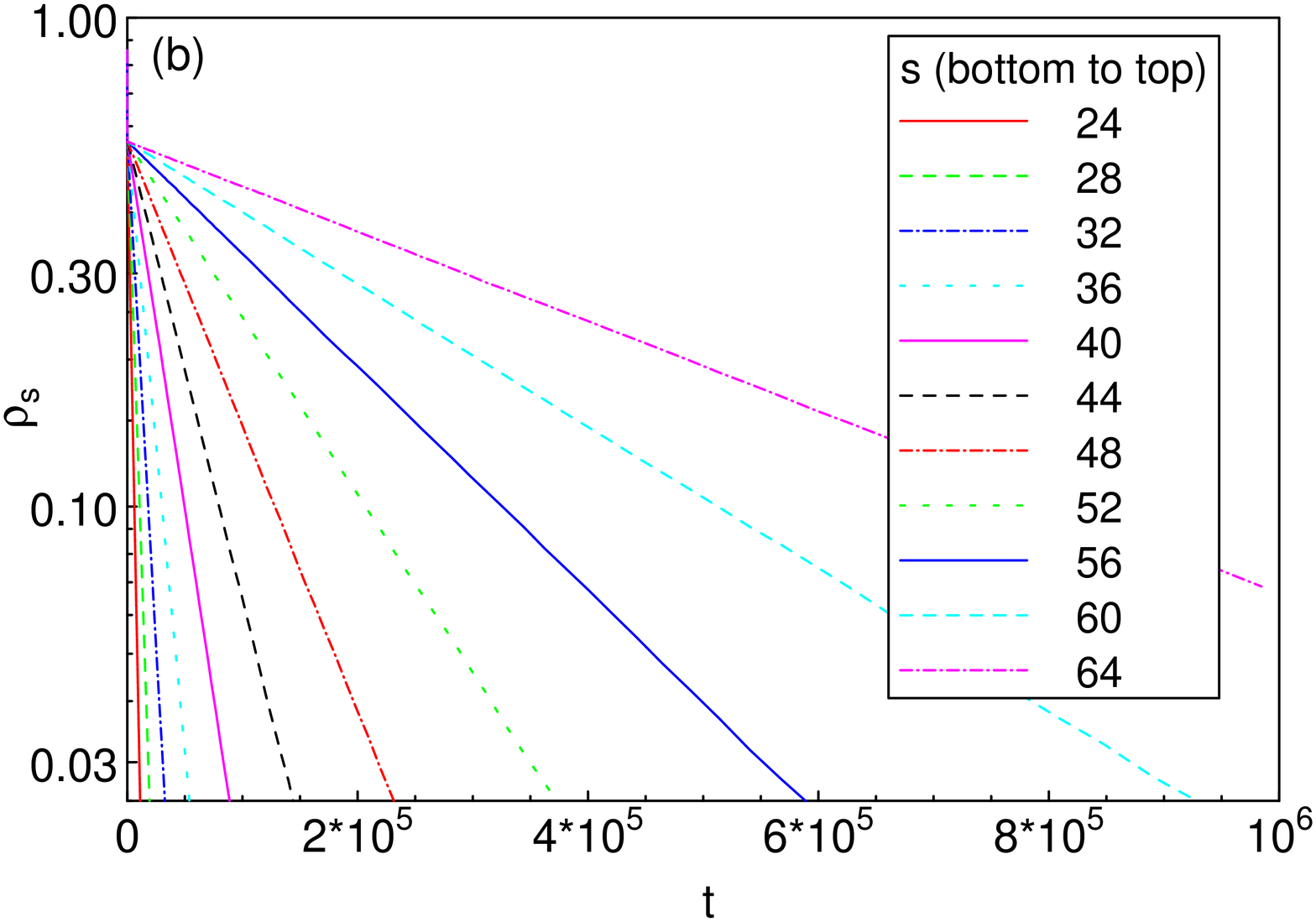}
\caption{(Color online:) Contact process on one-dimensional clusters of size $s$, starting from
a fully active lattice at $\lambda=3.8, \mu=1$ which is in the active phase. (a) Double-logarithmic plot
of density vs. time showing the two-stage time-evolution via a metastable state.
(b) Log-linear plot demonstrating that the
long-time decay is exponential. All data are averages over $10^5$ independent runs.}
\label{fig:decay_cp_1d}
\end{figure}

A characteristic set of results is shown in
Fig.\ \ref{fig:decay_cp_1d}.
It shows the time evolution of the contact process on several one-dimensional clusters of
different size $s$, starting from  a fully active lattice. The infection rate
$\lambda=3.8$ (we set $\mu=1$) puts the clusters (locally) in the ordered phase, i.e., it
is supercritical,
 since the critical value in one dimension is $\lambda_c=3.298$. All data
are averages over $10^5$ independent trials. The double-logarithmic plot of density $\rho_s$
vs. time $t$ in Fig.\ \ref{fig:decay_cp_1d}a clearly shows the two-stage time evolution
consisting of a rapid initial decay (independent of cluster size) towards a metastable
state followed by a long-time decay towards the absorbing state which becomes slower with
increasing cluster size. Replotting the data in log-linear form in Fig.\
\ref{fig:decay_cp_1d}b confirms that the long-time decay is exponential, as predicted
in (\ref{eq:cluster-decay}).

\begin{figure}
\includegraphics[width=8cm,clip]{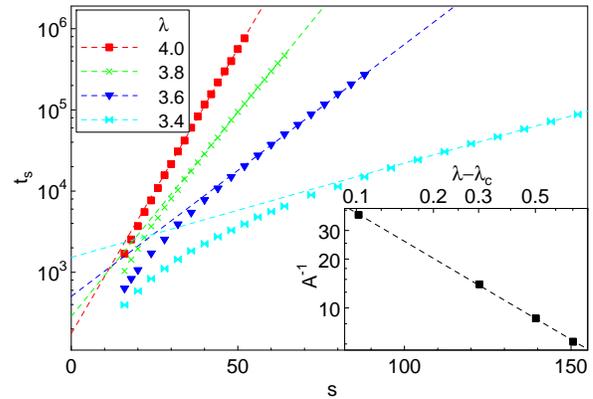}
\caption{(Color online:) Lifetime $t_s$ as a function of cluster size $s$ for different
          values of the infection rate $\lambda$. The other parameters are as in Fig.\
          \ref{fig:decay_cp_1d}. The dashed lines are fits of the large-$s$ behavior to
          the exponential dependence (\ref{eq:cluster-lifetime}). Inset: Correlation volume
          $A^{-1}$ as a function of the distance from bulk criticality. The dashed line is a
          power-law fit.}
\label{fig:lifetime-CP-1d}
\end{figure}

The lifetime $t_s$ of the contact process on the cluster can be determined by fitting
the asymptotic part of the $\rho_s(t)$ curve to (\ref{eq:cluster-decay}). Figure
\ref{fig:lifetime-CP-1d} shows the lifetime as a function of cluster size $s$ for four
different values of the infection rate $\lambda$.
Clearly, for sufficiently large clusters, the lifetime depends exponentially on the
cluster size, as predicted in  (\ref{eq:cluster-lifetime}). (The data for $\lambda=3.4$
which is very close to the bulk critical point of $\lambda_c=3.298$ have not fully
reached the asymptotic regime as can be seen from the remaining slight curvature of the
plot.) By fitting the large-$s$ behavior of the lifetime curves to the exponential law
 (\ref{eq:cluster-lifetime}), we obtain an estimate of the inverse correlation volume
$A$. The inset of Fig.\ \ref{fig:lifetime-CP-1d} shows this correlation volume as a
function of the distance from the bulk critical point. In accordance with (\ref{eq:Alambda})
it behaves as a power law. The exponent value of approximately 0.95 is in reasonable agreement
with the prediction $\nu = 1.097$ for our one-dimensional clusters.

We have performed analogous simulations for various sets of two-dimensional clusters as
well as finite-size diluted lattices. In all cases, the Monte-Carlo results confirm the
phenomenological theory summarized in eqs.\ (\ref{eq:cluster-decay}),
(\ref{eq:cluster-lifetime}), and (\ref{eq:Alambda}).

%%%%%%%%%%%%%%%%%%%%%%%%%%%%%%%%%%%%%%%%%%%%%%%%%%%%%%
\subsection{Steady-state density and density decay }
\label{subsec:density}
%%%%%%%%%%%%%%%%%%%%%%%%%%%%%%%%%%%%%%%%%%%%%%%%%%%%%%

We now consider the full problem, the contact process on a diluted lattice close to the
percolation threshold. To obtain observables of the entire system, we must sum over all
percolation clusters.

Let us start by analyzing static quantities such as the steady state density $\rho_{\rm st}$
of active sites (the order parameter of the nonequilibrium transition) and the spatial correlation length
$\xi_\perp$. Finite-size percolation clusters do not contribute to the steady-state density
because the contact process eventually decays into the absorbing inactive state on any
finite-size cluster. A steady-state density can thus exist only on the infinite
percolation cluster for $p<p_c$. For $\lambda>\lambda_*$, the infinite cluster is supercritical, i.e.,
a finite fraction of its sites is active.
Thus, the total steady-state density is proportional to the number of sites in the
infinite cluster,
\begin{equation}
\rho_{\rm st} \sim P_\infty(p) \sim  \left\{
\begin{array}{cc}  |p-p_c|^{\beta_c} & ~~ (p<p_c) \\
                       0             & ~~ (p>p_c)
                       \end{array} \right.~.
\label{eq:steady-state-density}
\end{equation}
Consequently, the order parameter exponent $\beta$ of the nonequilibrium transition is
identical to the corresponding exponent $\beta_c$ of the lattice percolation problem.

The (average) spatial correlation length $\xi_\perp$ of the nonequilibrium process can be
found using a similar argument. On the one hand, the spatial correlations of the contact process cannot
extend beyond the connectedness length $\xi_c$ of the underlying diluted lattice because
different percolation clusters are completely decoupled. This implies $\xi_\perp \lesssim
\xi_c$. On the other hand, for $\lambda>\lambda_*$, all sites on the same percolation
cluster are strongly correlated in space, implying $\xi_\perp \gtrsim \xi_c$. We
therefore conclude
\begin{equation}
\xi_\perp \approx \xi_c~,
\label{eq:spatial-correlation-length}
\end{equation}
and the correlation length exponent $\nu_\perp$ is also identical to its lattice percolation
counterpart $\nu_c$.

We now turn to the dynamics of the nonequilibrium transition across the percolation
threshold. In order to find the time evolution of the total density of active sites
(starting from a completely active lattice), we sum over all percolation clusters
by combining the cluster size distribution (\ref{eq:percscaling}) with the
single-cluster time evolution (\ref{eq:cluster-decay}). The total density is thus given
by
\begin{eqnarray}
\rho(t,\Delta) &=&   \int ds \, s \, n_s(\Delta)\, \rho_s(t) \nonumber \\
              &\sim& \int ds \, s \, n_s(\Delta)\, \exp[-t/t_s(s)]
\label{eq:total-density}
\end{eqnarray}
In the following, we evaluate this integral at the transition as well as in the
active and inactive phases.

Right at the percolation threshold, the scaling function in the cluster size distribution
(\ref{eq:percscaling}) is a constant, $f(0)$,  and (\ref{eq:total-density}) simplifies to
\begin{equation}
\rho(t,0) \sim \int ds ~s^{1-\tau_c} \exp[-(t/t_0e^{As})]~.
\label{eq:total-density-critical-integral}
\end{equation}
To estimate this integral, we note that only sufficiently large clusters, with a minimum
size of $s_{\rm min}(t)= A^{-1} \ln(t/t_0)$, contribute to the total density at time $t$,
\begin{equation}
\rho(t,0) \sim \int_{s_{\rm min}}^\infty ds \, s^{1-\tau_c} \sim s_{\rm min}^{2-\tau_c}~.
\label{eq:total-density-critical-estimate}
\end{equation}
The leading long-time dependence of the total density right at the percolation threshold
thus takes the unusual logarithmic form
\begin{equation}
\rho(t,0) \sim [\ln(t/t_0)]^{-\bar\delta} ~,
\label{eq:total-density-critical-result}
\end{equation}
again reflecting the activated dynamical scaling,
with the critical exponent given by $\bar\delta=\tau_c-2=\beta_c/(\nu_c D_f)$.

In the disconnected, inactive phase ($p>p_c$) we need to use expression
(\ref{eq:scaling-function-disconnected}) for the scaling function of the cluster size
distribution. The resulting integral for the time evolution of the density reads
\begin{equation}
\rho(t,\Delta) \sim  \int ds s^{1-\tau_c} \exp [-B_1 s \Delta^{1/\sigma_c} -(t/t_0 e^{A s})].
\label{eq:total-density-inactive-integral}
\end{equation}
For long times, the leading behavior of the integral can be calculated using the
saddle-point method. Minimizing the exponent of the integrand
shows that the main contribution at time $t$
to the integral (\ref{eq:total-density-inactive-integral}) comes from clusters of size
$s_0 = -A^{-1} \ln[B_1\Delta^{1/\sigma_c}t_0/(At)]$. Inserting this into the integrand
results in a power-law density decay
\begin{equation}
\rho(t,\Delta) \sim (t/t_0)^{-d/z'} \qquad (p>p_c)~.
\label{eq:total-density-inactive-result}
\end{equation}
The nonuniversal exponent $z'$ is given by  $z' =(
Ad/B_1)\Delta^{-1/\sigma_c} \sim \xi_\perp^{D_f}$, i.e.,
it diverges at the critical point $p=p_c$.

In the percolating, active phase ($p<p_c$), the infinite percolation cluster contributes
a nonzero steady state density $\rho_{\rm st}(\Delta)$ given by
(\ref{eq:steady-state-density}). However, the long-time approach of the
density towards this value is determined by the slow decay of the metastable states of
large finite-size percolation clusters. To estimate their contribution, we must
use the expression (\ref{eq:scaling-function-connected}) for the scaling function of the cluster size
distribution. The resulting integral now reads
\begin{eqnarray}
\rho(t,\Delta) -\rho_{st}(\Delta) &\sim& \int ds \,s^{1-\tau_c}
\exp\left[-(B_2 s |\Delta|^{1/\sigma_c})^{1-1/d} \right. \nonumber \\&&-
\left. (t/t_0 e^{As})\right]~.
\label{eq:total-density-active-integral}
\end{eqnarray}
We again apply the saddle-point method to find the leading low-time behavior of this
integral. Minimizing the exponent shows the main contribution coming from clusters of
size $s_0 = -A^{-1} \ln[B_2 |\Delta|^{1/\sigma_c}(d-1)/(Atd)]$. By inserting this into
the integrand, we find a nonexponential density decay of the form
\begin{equation}
\rho(t,\Delta)-\rho_{\rm st}(\Delta) \sim e^{ -\left[(d/z'') \ln(t/t_0) \right]^{1-1/d}}  \qquad
(p<p_c)~.
\label{eq:total-density-active-result}
\end{equation}
Here, $z''= (A d/B_2) |\Delta|^{-1/\sigma_c} \sim \xi_\perp^{D_f}$ is another
nonuniversal exponent which diverges at the critical point.

The slow nonexponential relaxation of the total density on both sides of the actual
transition as given in (\ref{eq:total-density-inactive-result}) and (\ref{eq:total-density-active-result})
is characteristic of a Griffiths phase \cite{Griffiths69} in the contact process
\cite{Noest88}. It is brought about by the competition between the exponentially
decreasing probability for finding a large percolation cluster off criticality and the
exponentially increasing lifetime of such a cluster. Note that time $t$ and spatial
correlation length $\xi_\perp$ enter the off-critical decay laws
(\ref{eq:total-density-inactive-result}) and (\ref{eq:total-density-active-result}) in
terms of the combination $\ln(t/t_0)/ \xi_\perp^{D_f}$ again reflecting the activated
character of the dynamical scaling.

%%%%%%%%%%%%%%%%%%%%%%%%%%%%%%%%%%%%%%%%%%%%%%%%%%
\subsection{Spreading from a single seed}
%%%%%%%%%%%%%%%%%%%%%%%%%%%%%%%%%%%%%%%%%%%%%%%%%%

After having discussed the time evolution of the density starting from a completely
infected lattice, we now consider the survival probability $P_s(t)$ for runs starting
from a single random seed site. To estimate $P_s(t)$, we note that the probability of a
random seed site to belong to a cluster of size $s$ is given by $s\,n_s(\Delta)$. The
activity of the contact process is confined to this seed cluster. Following the arguments
leading to (\ref{eq:cluster-decay}), the probability that this cluster survives is
proportional to $\exp(-t/t_s)$. The average survival probability at time $t$ can thus be
written as a sum over all possible seed clusters,
\begin{equation}
P_s(t,\Delta) \sim \int ds \, s \, n_s(\Delta)\, \exp[-t/t_s(s)]~.
\label{eq:survival-probability-integral}
\end{equation}
This is exactly the same integral as the one governing the density decay
(\ref{eq:total-density}). We conclude that the time dependence of the survival
probability for runs starting from a single seed is identical to the time evolution  of
the density when starting from a fully infected lattice, as is expected for the contact
process under very general conditions (see, e.g., Ref.\ \cite{Hinrichsen00}).

To determine the (average) total number $N(t)$ of active sites in a cloud spreading from a single
seed, we observe that a supercritical cloud initially grows ballistically.
This means its radius grows linearly with time, and the number of
active sites follows a power law.  This ballistic growth stops when the number of active sites
is of the order of the
cluster size $s$. After that, the number of active sites stays approximately
constant. The number $N_s(t)$ of active sites on a percolation cluster of size $s$ is
thus given by
\begin{equation}
N_s(t) \sim \left \{
\begin{array}{cc}
(t/t_0)^{D_f}  & \quad (t<t_i(s))\\
s              & \quad (t>t_i(s))
\end{array}
\right.
\label{eq:Ns}
\end{equation}
where $t_i(s) \sim R_s(s) \sim t_0 s^{1/D_f}$ is the saturation time of this cluster.
Note that $N_s$ decays to zero only after the much longer cluster lifetime $t_s(s)=t_0
\exp[{A(\lambda)s}]$ given in (\ref{eq:cluster-lifetime}).

We now average over all possible positions of the seed site as in
(\ref{eq:survival-probability-integral}). This yields
\begin{equation}
N(t,\Delta) \sim \int_{s_{\rm min}}^\infty  ds \, s \, n_s(\Delta) \, N_s(t)
\label{eq:total-N}
\end{equation}
with $s_{\rm min}\sim A^{-1} \ln(t/t_0)$. At criticality, this integral is easily
evaluated, giving
\begin{equation}
N(t,0) \sim t^{D_f(3-\tau_c)}= t^{\gamma_c/\nu_c}~.
\label{eq:N-result}
\end{equation}
The lower bound of the integral (i.e., the logarithmically slow long-time decay of the
clusters) produces a subleading correction only. Consequently, we arrive at the somewhat
surprising conclusion that the initial spreading follows a power-law and is thus much
faster than the long-time density decay. In contrast, at the infinite-randomness critical
point governing the generic ($p<p_c$) transition, both the initial spreading and the
long-time decay follow logarithmic laws
\cite{HooyberghsIgloiVanderzande03,HooyberghsIgloiVanderzande04,VojtaDickison05,VojtaFarquharMast09}.
Note that a similar situation occurs at the percolation quantum phase transition in the
diluted transverse-field Ising model \cite{SenthilSachdev96} where the
temperature-dependence of the correlation length does not follow the naively expected
logarithmic law.

%%%%%%%%%%%%%%%%%%%%%%%%%%%%%%%%%%%%%%%%%%%%%%%%%%%%%%
\subsection{External source field}
\label{subsec:source}
%%%%%%%%%%%%%%%%%%%%%%%%%%%%%%%%%%%%%%%%%%%%%%%%%%%%%%

In this subsection we discuss the effects of spontaneous activity creation on
our nonequilibrium phase transition. Specifically, in addition to healing and infection,
we now consider a third process by which an inactive site can spontaneously turn into
an active site at rate $h$. This rate plays the role of an external ``source field'' conjugate
to the order parameter.

To find the steady state density in the presence of such a source field, we first
consider a single percolation cluster. As before, we are interested in the supercritical regime
$\lambda>\lambda_*$. At any given time $t$, a cluster of size $s$ will be active (on
average), if at least one of the $s$ sites has spontaneously become active within
one lifetime $t_s(s)=t_0 e^{As}$ before $t$, i.e., in the interval $[t-t_s(s),t]$. For a
small external field $h$, the average number of active sites created on a cluster of size
s is $M_s(h) = hst_s(s) = hst_0 e^{As}$. This linear response expression is valid as
long as $M_s \ll s$. The probability $w_s(h)$ for a cluster of size $s$ to be active
in the steady state is thus given by
\begin{equation}
w_s(h) \approx \left \{
\begin{array}{cc}
M_s(h) & \quad (M_s(h)<1)\\
1  & \quad  (M_s(h)>1)
\end{array}
\right. ~.
\label{eq:field-probability}
\end{equation}

Turning to the full lattice, the total steady state density is obtained by summing over
all clusters
\begin{equation}
\rho_{\rm st}(h,\Delta) \sim \int ds \, s \, n_s(\Delta) \min[1,M_s(h)]~.
\label{eq:density-field-integral}
\end{equation}
This integral can be evaluated along the same lines as the corresponding integral
(\ref{eq:total-density}) for the time-evolution of the zero-field density.
For small fields $h$, we obtain
\begin{eqnarray}
\rho_{\rm st} (h,0)  &\sim& [\ln(h_0/h)]^{-\bar\delta}~ \qquad\qquad (p=p_c), \label{eq:rho-h-critical}\\
\rho_{\rm st} (h,\Delta) &\sim& (h/h_0)^{d/z'} ~\qquad\qquad (p>p_c), \\
\delta\rho_{\rm st} (h,\Delta) &\sim& e^{\left[ (d/z'') \ln(h/h_0)
\right]^{1-1/d}} \quad (p<p_c)~,
\end{eqnarray}
where $\delta\rho_{\rm st} (h,\Delta)=\rho_{\rm st} (h,\Delta)-\rho_{\rm st} (0,\Delta)$ is the excess
density due to the field in the active phase and $h_0 = 1/t_0$. At criticality, $p=p_c$, the relation between density $\rho_{\rm st}$ and field $h$ is
logarithmic because the field represents a rate (inverse time) and the dynamical scaling is activated.
Off criticality, we find strong Griffiths singularities analogous to those in the time-dependence of the density.
The exponents $z'$ and $z''$ take the same values as calculated after eqs.\ (\ref{eq:total-density-inactive-result})
and (\ref{eq:total-density-active-result}), respectively.

%%%%%%%%%%%%%%%%%%%%%%%%%%%%%%%%%%%%%%%%%%%%%%%%%%%%%%
\subsection{Scaling theory}
%%%%%%%%%%%%%%%%%%%%%%%%%%%%%%%%%%%%%%%%%%%%%%%%%%%%%%

In Sections \ref{subsec:density} and \ref{subsec:source}, we have determined the critical
behavior of the density of active sites by explicitly averaging the single cluster
dynamics over all percolation clusters. The same results can also be obtained from
writing down a general scaling theory of the density for the case of activated dynamical
scaling \cite{VojtaDickison05,VojtaFarquharMast09}.

According to (\ref{eq:steady-state-density}), in the active phase, the density is
proportional to the number of sites in the infinite percolation cluster. Its scale
dimension must therefore be identical to the scale dimension of $P_\infty$ which is
$\beta_c/\nu_c$. Time must enter the theory via the scaling combination
$\ln(t/t_0)b^\psi$ with the tunneling exponent given by $\psi=D_f$ and $b$ an arbitrary
length scale factor. This scaling combination reflects the activated dynamical scaling,
i.e., the exponential relation (\ref{eq:activated-scaling}) between length and time
scales. Finally, the source field $h$, being a rate, scales like inverse time. This leads
to the following scaling theory of the density,
\begin{eqnarray}
 \rho[\Delta,\ln(t/t_0),\ln(h_0/h)] = \hspace*{4cm} \nonumber \\
=b^{\beta_c/\nu_c}\rho[\Delta b^{-1/\nu_c}, \ln(t/t_0) b^\psi,
\ln(h_0/h) b^\psi]~
\label{eq:density-scaling}
\end{eqnarray}
This scaling theory is compatible with all our explicit results which can be rederived by
setting the arbitrary scale factor $b$ to the appropriate values.

%%%%%%%%%%%%%%%%%%%%%%%%%%%%%%%%%%%%%%%%%%%%%%%%%%%%%%%%%%%%%%%%%%%%%%%%%%%%%%%%%
\section{Generality of the activated scaling scenario}
\label{sec:generality}
%%%%%%%%%%%%%%%%%%%%%%%%%%%%%%%%%%%%%%%%%%%%%%%%%%%%%%%%%%%%%%%%%%%%%%%%%%%%%%%%%

In Section \ref{sec:theory}, we have developed a theory for the nonequilibrium phase
transition of the simple contact process across the lattice percolation threshold and
found it to be characterized by unconventional activated dynamical scaling.
In the present section, we investigate how general this
exotic behavior is for absorbing state transitions by considering the generalized contact
process with several absorbing states.

This is a particularly interesting question because the generic transitions ($p<p_c$)
of the diluted simple and generalized contact processes appear to behave differently.
The generic transition in the simple contact process has been shown to be of
infinite-randomness type with activated dynamical scaling using both a strong-disorder
renormalization group \cite{HooyberghsIgloiVanderzande03,HooyberghsIgloiVanderzande04}
and Monte-Carlo simulations \cite{VojtaDickison05,VojtaFarquharMast09}. In contrast,
the strong-disorder renormalization group treatment of the disordered generalized contact
process \cite{HooyberghsIgloiVanderzande04} suggests more conventional behavior, even
though the ultimate fate of the transition could not be determined.

To address the same question for our transition across the lattice percolation threshold,
we note that any difference between the simple and the generalized contact processes must
stem from the single-cluster dynamics because the underlying lattice is identical. In the
following we therefore first give heuristic arguments for the single-cluster dynamics of the
supercritical generalized contact process and then verify them by Monte-Carlo simulations.

\begin{figure}
\includegraphics[width=6.5cm,clip]{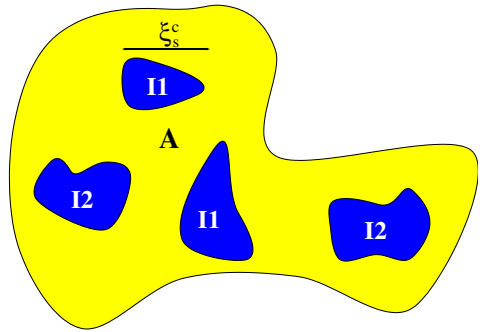}
\caption{(Color online:) Schematic of the metastable state of the supercritical
         generalized contact process
         with two inactive states on a single percolation cluster.
         A denotes the active state, and I$_1$ and I$_2$ are the inactive states.
         $\xi_s^c$ is the connected
         correlation length of the density fluctuations \emph{on} the cluster.}
\label{fig:gcp_cluster}
\end{figure}

If the percolation cluster is locally in the active phase ($\lambda>\lambda_*$), the
density time evolution, starting from a fully active lattice, proceeds in two stages,
analogously to the simple contact process. There is a rapid initial decay to a metastable
state with a nonzero
density of active sites and finite-size islands of each of the inactive phases
(see Fig.\ \ref{fig:gcp_cluster}).
For this metastable state to decay into one of the $n$ absorbing configurations, all
sites must go into \emph{the same} inactive state which requires a rare large density
fluctuation. Let us assume for definiteness that the decay is into the I$_1$ state. The
main difference to the simple contact process considered in Sec.\
\ref{subsec:single-cluster} is that sites that are in inactive states I$_2 \ldots$ I$_n$
cannot directly decay into I$_1$. This means, each of the inactive islands in states I$_2
\ldots$ I$_n$ first needs to be ``eaten'' by the active regions before the entire cluster
can decay into the I$_1$ state. This can only happen via infection from the boundary of
the inactive island and is thus a slow process. However, since the characteristic size of
the inactive islands in the metastable state is finite (it is given by the connected
density correlation length $\xi_s^c$ on the cluster), this process happens with a nonzero
rate that is independent of the size $s$ of the underlying percolation cluster (for
sufficiently large $s$).

The decay of the metastable state into one of the absorbing states is therefore brought
about by the rare collective decay of a large number of \emph{independent} correlation
volumes just as in the simple contact process. As a result, the lifetime $t_s(s)$ depends
exponentially on the number of involved correlation volumes, i.e., it depends
exponentially on the cluster size $s$. We thus find that the long-time density decay of
the generalized contact process on a single large percolation cluster is governed by
the same equations (\ref{eq:cluster-decay}) and (\ref{eq:cluster-lifetime}) as the decay
of the simple contact process.

\begin{figure}
\includegraphics[width=8cm]{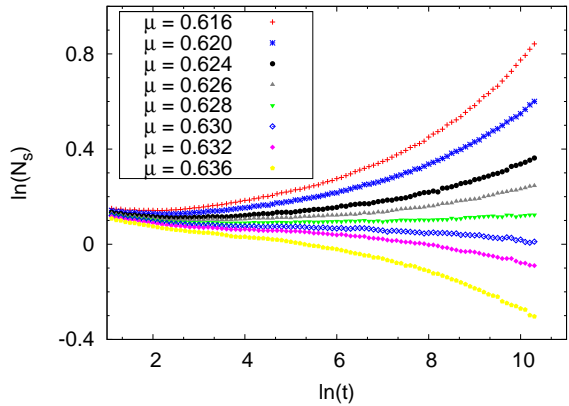}
\caption{(Color online:) Bulk phase transition of the generalized contact process with two
absorbing states in $d=1$ measured via spreading from a single seed: Number $N$ of active
sites vs. time $t$ for different healing rates $\mu$. The infection and boundary activation
rates are fixed, $\lambda=\sigma=1$, and the data are averages over $10^6$ runs. The critical
point appears to be close to $\mu=0.628$ in agreement with \cite{Hinrichsen97}.}
\label{fig:spreading-gcp}
\end{figure}

To verify these phenomenological arguments, we have performed large-scale Monte-Carlo
simulations of the generalized contact process with two and three absorbing states
on clean and disordered one-dimensional and two-dimensional lattices. In all cases, we have
first performed bulk simulations (spreading from a single seed) to find the bulk critical
point. An example is shown in Fig.\ \ref{fig:spreading-gcp}, details of the bulk critical
behavior will be reported elsewhere.

\begin{figure}
\includegraphics[width=8cm]{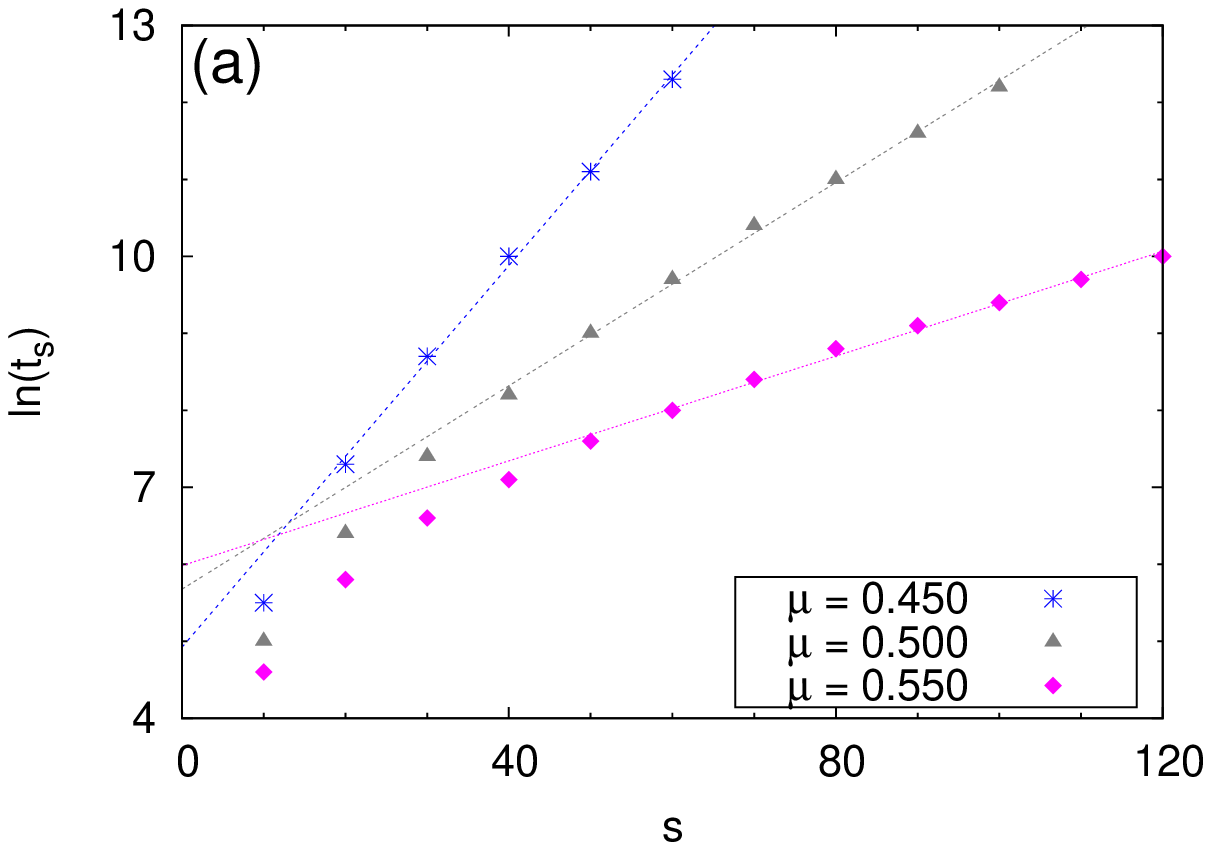}
\includegraphics[width=8cm]{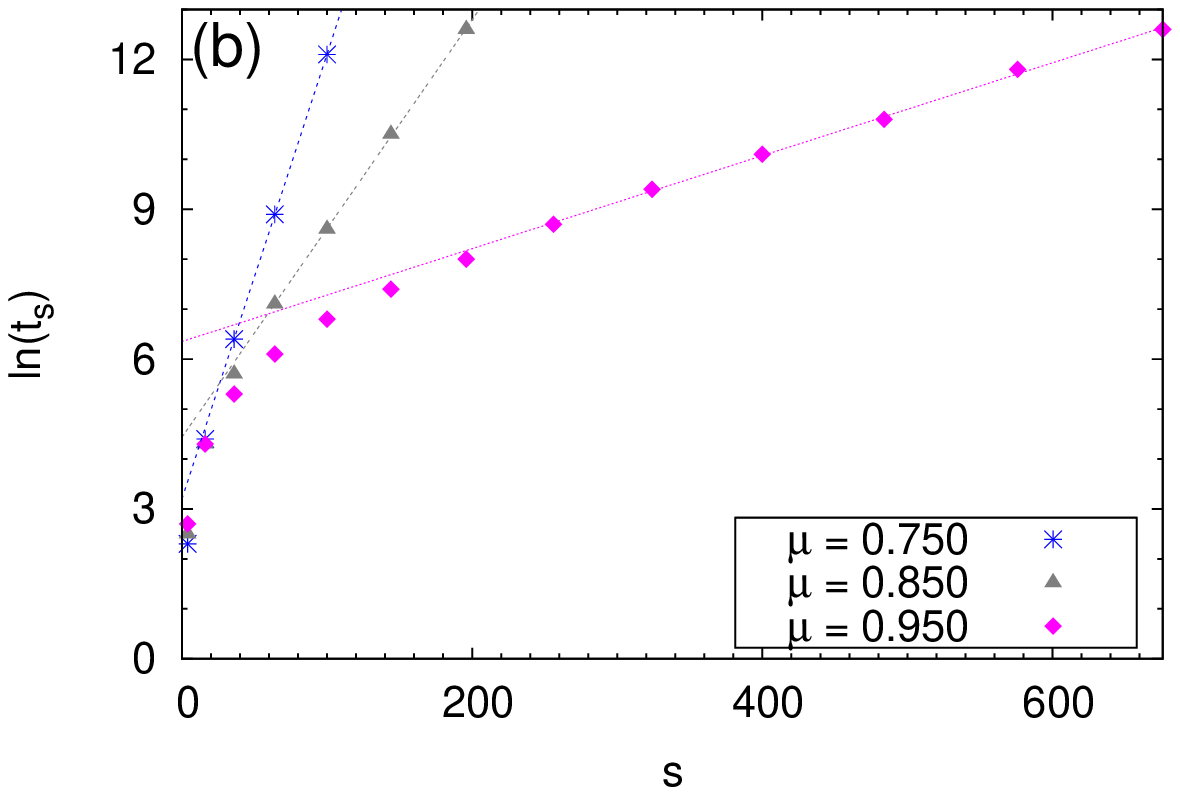}
\caption{(Color online:) Lifetime $t_s$ as a function of cluster size $s$ for the generalized
contact process with two inactive states at different values of the healing rate $\mu$.
The infection and boundary activation rates are fixed, $\lambda=\sigma=1$, and the data are
averages over $10^6$ runs. (a) $d=1$ where the bulk system has a transition, see Fig.\
\ref{fig:spreading-gcp}. (b) $d=2$, where we do not find a bulk transition because the system
is always active \cite{Bibnote1}. The dashed lines are fits of the large-$s$ behaviors to
the exponential law (\ref{eq:cluster-lifetime}). }
\label{fig:decay-gcp}
\end{figure}

After having determined the critical point, if any, we have selected several parameter
sets in the bulk active phase and studied the long-time density decay of the generalized
contact process on finite size clusters. As expected, the decay proceeds via the two
stages discussed above. As in Sec.\ \ref{subsec:single-cluster}, we extract the lifetime
$t_s$ from the slow exponential long-time part of the decay. Two characteristic sets of
results are shown in Fig.\ \ref{fig:decay-gcp}.
The figure confirms that the lifetime of the generalized contact process on a finite-size
cluster depends exponentially on the number of sites in the cluster, as given in
(\ref{eq:cluster-lifetime}). We have obtained analogous results for all cases investigated,
verifying the phenomenological theory given above.

Because the long-time dynamics of the generalized contact process on a single supercritical
cluster follows the same behavior (\ref{eq:cluster-decay}) and
(\ref{eq:cluster-lifetime}) as that of the simple contact process,
we conclude that its nonequilibrium transition across the percolation threshold will also be
governed by the theory developed on Sec.\ \ref{sec:theory}. In other words, the lattice
percolation transitions of the simple and generalized contact processes belong to the same
universality class, irrespective of the number $n$ of absorbing states.

%%%%%%%%%%%%%%%%%%%%%%%%%%%%%%%%%%%%%%%%%%%%%%%%%%%%%%%%%%%%%%%%%%%%%%%%%%%%%%%%%
\section{Conclusions}
\label{sec:conclusions}
%%%%%%%%%%%%%%%%%%%%%%%%%%%%%%%%%%%%%%%%%%%%%%%%%%%%%%%%%%%%%%%%%%%%%%%%%%%%%%%%%

In this final section of the paper, we first summarize our results, discuss their
generality,
and relate them to the behavior of certain quantum phase transitions on diluted lattices. We then
compare the recently found infinite-randomness critical point at the generic transition
($p<p_c$) to the behavior at our lattice percolation transition. Finally, we relate our
findings to a general classification of phase transitions with quenched spatial disorder
\cite{Vojta06}.

To summarize, we have investigated absorbing state phase transitions on randomly diluted
lattices, taking the simple and generalized contact processes as examples. We have
focused on the nonequilibrium phase transition across the lattice percolation threshold
and shown that it can be understood by combining the time evolution of the supercritical
nonequilibrium process on a finite-size cluster with results from classical lattice
percolation theory. The interplay between geometric criticality and dynamic fluctuations
at this transition leads to a novel universality class. It is characterized by ultraslow
activated (i.e., exponential) rather than power-law dynamical scaling and accompanied by
a nonexponential decay in the Griffiths regions.
All critical exponents of the nonequilibrium phase transition can be expressed in terms
of the classical lattice percolation exponents. Their values are known exactly in two
space dimensions and with good numerical accuracy in three space dimensions; they are
summarized in Table \ref{table:exponents}.
\begin{table}[tbp]
\caption{Critical exponents of the nonequilibrium phase transition across the percolation
threshold in two and three space dimensions. } \label{table:exponents}
\begin{ruledtabular}
\begin{tabular}{l c c}
Exponent    & $~d=2~$ & $~d=3~$ \\
\hline
$\beta  =\beta_c$                   & 5/36 & 0.417  \\
$\nu    =\nu_c$                     &  4/3 & 0.875  \\
$\psi   =D_f=d-\beta_c/\nu_c$       & 91/48& 2.523  \\
$\bar\delta=\beta_c/(\nu_c D_f) $   & 5/91 & 0.188  \\
\end{tabular}%
\end{ruledtabular}
\end{table}
Thus, our transition in $d=2$ provides one of the few examples of a nonequilibrium phase
transition with exactly known critical exponents.

The logarithmically slow dynamics (\ref{eq:total-density-critical-result}), (\ref{eq:rho-h-critical})
at criticality together with the small value of the exponent $\bar\delta$ make a
numerical verification of our theory by simulations of the full diluted lattice a very
costly proposition. The results of recent Monte-Carlo simulations in two dimensions
\cite{VojtaFarquharMast09} at $p=p_c$ are compatible with our theory but not yet sufficient
to be considered a quantitative verification. This remains a task for the future.

The unconventional critical behavior of our nonequilibrium phase transition at $p=p_c$ is
the direct result of combining the power-law spectrum (\ref{eq:percscaling}) of cluster
sizes with the exponential relation (\ref{eq:activated-scaling}) between length and time
scales. We therefore expect other equilibrium or nonequilibrium systems that share these
two characteristics to display similar critical behavior at the lattice percolation
transition. One prototypical example is the transverse-field Ising model on a diluted
lattice. In this system, the quantum-mechanical energy gap (which represents an inverse
time) of a cluster decreases exponentially with the cluster size. Consequently, the
critical behavior of the diluted transverse-field Ising model across the lattice
percolation threshold is very similar to the one found in this paper
\cite{SenthilSachdev96}. Other candidates are magnetic quantum phase transitions in
metallic systems or certain superconductor-metal quantum phase transitions
\cite{VojtaSchmalian05,HoyosKotabageVojta07,DRMS08,VojtaKotabageHoyos09}, even though a
pure percolation scenario may be hard to realize in metallic systems.

Our work has focused on the nonequilibrium phase transition across the lattice
percolation threshold. It is instructive to compare its critical behavior to that of the
generic transition occurring for $p<p_c$ (see Fig.\ \ref{fig:pds}). Hooyberghs et al.
\cite{HooyberghsIgloiVanderzande03,HooyberghsIgloiVanderzande04} applied a strong
disorder renormalization group to the one-dimensional disordered contact process. They
found an exotic infinite-randomness critical point in the universality class of the
random-transverse field Ising model (which likely governs the transition for any disorder
strength \cite{Hoyos08}). The same analogy is expected to hold in two space
dimensions. Recently, these predictions were confirmed by large scale Monte-Carlo
simulations \cite{VojtaDickison05,VojtaFarquharMast09}. Our nonequilibrium transition
across the lattice percolation threshold shares some characteristics with these
infinite-randomness critical points,  in particular, the activated dynamical scaling
which leads to a logarithmically slow density decay at criticality.

However, the generic and percolation transitions are in different universality classes
with different critical exponent values.  Moreover, the initial spreading from a single
seed is qualitatively different (logarithmically slow at the generic infinite-randomness
critical point but of power-law type at our percolation transition). Finally, at the
percolation transition the simple and generalized contact processes are in the same
universality class while this does not seem to be the case for the generic transition
\cite{HooyberghsIgloiVanderzande04}.

The results of this paper are in agreement with a recent general classification of phase
transitions with quenched spatial disorder and short-range interactions
\cite{VojtaSchmalian05,Vojta06}. It is based on the effective dimensionality $d_{\rm
eff}$ of the droplets or clusters. Three classes need to be distinguished: (a) If the
clusters are below the lower critical dimension of the problem, $d_{\rm eff} < d_c^-$,
the critical behavior is conventional (power-law scaling and exponentially weak Griffiths
effects). This is the case for most classical equilibrium transitions.  (b) If $d_{\rm
eff} = d_c^-$, the dynamical scaling is activated and accompanied by strong Griffiths
effects. This case is realized at the nonequilibrium transition considered here as well
as the generic transition of the disordered contact process. It also applies to various
quantum phase transitions \cite{Fisher92,SenthilSachdev96,HoyosKotabageVojta07}. (c) If
$d_{\rm eff} > d_c^-$, a single supercritical cluster can undergo the phase transition
independently of the bulk system. This leads to the smearing of the global phase
transition; it occurs, e.g., in dissipative quantum magnets \cite{Vojta03a,HoyosVojta08}
or in the contact process with extended defects \cite{Vojta04}.

In conclusion, our work demonstrates that absorbing state transitions on percolating
lattices display unusual behavior. Interestingly, experimental verifications of the
theoretically predicted critical behavior at (clean) absorbing state transitions are
extremely rare \cite{Hinrichsen00b}. For instance, to the best of our knowledge, the only
complete verification of directed percolation scaling was found very recently in the
transition between two turbulent states in a liquid crystal \cite{TKCS07}. Our theory
suggests that unconventional disorder effects may be responsible for the surprising
absence of directed percolation scaling in at least some of the experiments.

%%%%%%%%%%%%%%%%%%%%%%%%%%%%%%%%%%%%%%%%%%%%%%%%%%%%%%%%%%%%%%%%%%%%%%%%%%%%%%%%%
\section*{Acknowledgements}
%%%%%%%%%%%%%%%%%%%%%%%%%%%%%%%%%%%%%%%%%%%%%%%%%%%%%%%%%%%%%%%%%%%%%%%%%%%%%%%%%

This work has been supported in part by the NSF under grant no. DMR-0339147, by Research
Corporation, and by the University of Missouri Research Board. We gratefully acknowledge
discussions with J. Hoyos as well the hospitality of the
Max-Planck-Institute for Physics of Complex Systems during part of this research.

\bibliographystyle{apsrev}
\bibliography{../00Bibtex/rareregions}

\begin{thebibliography}{49}
\bibitem[{Bib({\natexlab{a}})}]{Bibnote1}
\bibinfo{note}{For $d=2, n=2$, Hinrichsen \cite {Hinrichsen97} finds a
  mean-field transition while our own simulations suggest that the system
  always active. Since this difference is of no importance for the present
  paper, it will be addressed elsewhere.}
\bibitem[{Bib({\natexlab{b}})}]{Bibnote2}
\bibinfo{note}{We define $p$ as the fraction of sites removed rather than the
  fraction of sites present.}
\end{thebibliography}
\end{document}